\documentstyle[12pt]{article}
\newcommand{\figurebox}[2]{\mbox{\vbox to#2in{\hbox to #1in{\hfil}
\vfil}}}

\begin{document}
\renewcommand{\thefootnote}{\fnsymbol{footnote}}
                                        \begin{titlepage}
\begin{flushright}
hep-ph/9310316
\end{flushright}
\vskip0.8cm
\begin{center}
{\LARGE  ERRATUM TO:\\
Power corrections $1/Q^2$ to parton sum rules for deep
inelastic scattering from polarized targets
            \\ }
\vskip1cm
{\Large I. I. Balitsky}~$^\ast$ \\
\vskip0.2cm
        Physics Department, Penn State University,\\
       104 Davey Lab.,University Park, PA 16802, USA\\
\vskip0.5cm
 {\Large V. M.~Braun} $\footnote { On leave of absence from
St.Petersburg Nuclear
Physics Institute, 188350 Gatchina, Russia}$ \\
\vskip0.2cm
       Max-Planck-Institut f\"ur Physik   \\
       -- Werner-Heisenberg-Institut -- \\
        D--80805 Munich (Fed. Rep. Germany)\\
and\\
\vskip0.5cm
 {\Large  A. V.~Kolesnichenko}
 $\footnote { deceased  }$ \\
\vskip0.2cm
St.Petersburg Nuclear
Physics Institute, 188350 Gatchina, Russia \\
 \vspace{1.0cm}
{\em published in Physics Letters B242 (1990) 245 }
\vskip1cm
{\Large Abstract:\\}
\parbox{\textwidth}{
 We clarify conflicting results in the literature on
 coefficient
functions in front of higher twist operators contributing to the
parton sum rules for deep inelastic scattering from polarized
 targets. The necessary corrections do not affect our calculations
of matrix elements, but change final estimates of  the $\sim 1/Q^2$
contributions to  Bjorken and Ellis--Jaffe sum rules.
}
\end{center}
                                                \end{titlepage}

\newpage
\setcounter{equation} {13}
Recently, Ji and Unrau \cite{JU} have pointed out that coefficients
in front of higher twist contributions to the sum rules for
 polarized
deep inelastic scattering given in \cite{SV} contain errors.
We have checked the calculation of \cite{SV}, and indeed  have
found that the factors 8/3 in front of the kinematical
corrections given in eq.(51) of \cite{SV} must be replaced
by 4/3, in agreement with \cite{JU}. In addition, we have found
an overall sign error in the expressions for the moments of
$g_2(x)$ given in eq.(47) of  \cite{SV}.
The statement of \cite{JU} about the sign error in the
coefficient of the twist 4 operator is not correct.
The difference in sign between \cite{SV} and \cite{JU} is due
to different conventions for the $\gamma_5$-matrix and the
$\epsilon_{\alpha\beta\mu\nu}$-tensor. The conventions used
in \cite{SV,BBK} are taken from \cite{OKUN}.
We thank A.~Vainshtein for the correspondence on this point,
and understand that authors of \cite{SV} agree to the changes
specified above.

The calculations of power corrections in \cite{BBK} have used
the expressions given in \cite{SV}, and must be corrected,
respectively.
\begin{itemize}
\item
Throughout the paper, the coefficients in front of the
kinematical power corrections $\sim m_N^2$ acquire an additional
factor 1/2.
\item
In eq.(4) the sign of the second term (twist three)
    should be reversed. This induces the sign change in front
  of the second terms in eq.(16), eq.(17) and in the non-numbered
equation after eq.(5), which becomes
$$
-\frac{8}{9Q^2}\left (\langle\!\langle U \rangle\!\rangle -
\frac{1}{4} m_N^2 \langle\!\langle V \rangle\!\rangle\right ) =
 - \frac{\langle\!\langle O \rangle\!\rangle}{Q^2}
$$
\item
We thank D. Roberts for pointing to us an error in the
stability plot for $ \langle\!\langle U \rangle\!\rangle $
shown in Fig.2, which is due to the error in the computer
program. The correct value for the singlet matrix element
$ \langle\!\langle U \rangle\!\rangle  \simeq 0.05 \,GeV^2 $
is given in the preliminary publication of this paper \cite{BBK1}.
\item
We take this opportunity to correct misprints. The factor $m_N^2$ should
be added on the r.h.s. of eq.(7). In the sum rules in eq.(11)
the generic operator
$O$ stands for either $U$ or $m_N^2 V$.
The second formula in (16) gives the second moment of  $g^{p-n}_2$.
\end{itemize}
Our final values for the particular combination of twist 4 and twist 3
matrix elements defined above are
\begin{eqnarray}
\langle\!\langle O^{NS} \rangle\!\rangle = 0.09\pm 0.06\, GeV^2,
&&
\langle\!\langle O^{S} \rangle\!\rangle = 0.09\pm 0.06\, GeV^2.
\nonumber    \\
\langle\!\langle O^{p} \rangle\!\rangle = 0.09\pm 0.06 \,GeV^2,
&&
\langle\!\langle O^{n} \rangle\!\rangle = 0.0 \pm 0.03 \, GeV^2.
\nonumber
\end{eqnarray}
The  final eqs.(14),(15) should read
\begin{equation}
\int dx\,g_1^{p-n}(x,Q^2) =
\frac{1}{6}\left\{g_A[1-\frac{\alpha_s(Q^2)}{\pi}]
-\frac{(0.09\pm 0.06) GeV^2}{Q^2} \right\}
+\frac{2}{9}\frac{m_N^2}{Q^2}
\int dx\, x^2 g_1^{p-n}(x),
\end{equation}
\begin{equation}
\int dx\,g_1^{p+n}(x,Q^2) =
\frac{5}{18}\left\{g_A^S[1-\frac{\alpha_s(Q^2)}{\pi}]
-\frac{(0.09\pm 0.06) GeV^2}{Q^2} \right\}
+\frac{2}{9}\frac{m_N^2}{Q^2}
\int dx\, x^2 g_1^{p+n}(x).
\end{equation}
\medskip
We gratefully acknowledge the correspondence with X.~Ji, M.~Karliner,
D.~Roberts and A.~Vainshtein, whose interest initiated this Erratum.

\end{document}